# Electroluminescence from Strained Ge membranes and Implications for an Efficient Si-Compatible Laser

Donguk Nam,[1,*] David Sukhdeo,[1] Szu-Lin Cheng,[3] Arunanshu Roy,[1] Kevin Chih-Yao Huang,[1] Mark Brongersma,[2] Yoshio Nishi[1], and Krishna Saraswat[1]

[1]Department of Electrical Engineering, Stanford University, Stanford, CA 94305, USA
[2]Department of Materials Science and Engineering, Stanford University, Stanford, CA 94305, USA
[3]IBM T.J. Watson Research Center, 1101 Kitchawan Road, Yorktown Heights, NY 10598, USA
[*]dwnam@stanford.edu

**Abstract:** We demonstrate room-temperature electroluminescence (EL) from light-emitting diodes (LED) on highly strained germanium (Ge) membranes. An external stressor technique was employed to introduce a 0.76% bi-axial tensile strain in the active region of a vertical PN junction. Electrical measurements show an on-off ratio increase of one order of magnitude in membrane LEDs compared to bulk. The EL spectrum from the 0.76% strained Ge LED shows a 100nm redshift of the center wavelength because of the strain-induced direct band gap reduction. Finally, using tight-binding and FDTD simulations, we discuss the implications for highly efficient Ge lasers.

The use of Germanium (Ge) for on-chip optical interconnects has recently gained an increasing amount of interest because it can be monolithically integrated on silicon (Si) substrates [1-2]. In order to achieve a complete optical link, researchers have been extensively working on Si and Ge devices, including modulators, waveguides, and detectors [3-4]. Despite successful demonstrations of those devices, a Ge light source remains particularly challenging due to its indirect band gap. Fortunately, however, the energy difference between the direct Γ valley and the indirect L valley is only 136meV. Moreover, this difference can be reduced further by introducing tensile strain in Ge due to the different deformation potential parameters in the two valleys [5]. Recently, heavily n-type doped Ge under 0.2% tensile strain from thermal mismatch during Ge growth has recently shown direct band gap light emission attributed to the band filling effect and the direct band gap reduction [6-8]. Also, there has been much effort to improve light emission efficiency more by alloying with tin (Sn) and by mechanically introducing larger tensile strains, since both Sn and strain are known to reduce Ge's direct band gap energy faster than its indirect band gap energy [9-12]. Notably, recent experiments have shown that a 1.13% tensile strain can be permanently induced in a Ge membrane using an external stressor, and Ge photodetectors on these highly strained membranes were shown to exhibit an excellent responsivity at wavelengths well beyond 1.7μm [13].

In this letter, we present an LED on a highly strained Ge membrane. Electrical measurements show an increase in forward current by a factor of 3 with 0.76% bi-axial tensile strain in the active region of a vertical PN diode compared to 0.2% strain. With respect to optical properties, a 0.76% strained LED exhibits a 100nm red-shift in the EL spectrum compared to 0.2% strained case. To investigate the implications of this work, we propose a vertical-cavity surface-emitting laser (VCSEL) using a highly strained Ge membrane as the gain medium. Using finite-difference in time domain (FDTD) simulations and a tight-binding band structure model, we predict that the threshold current density in a 1% strained VCSEL will be reduced by more than a factor of 3 compared to a 0.2% strained case.

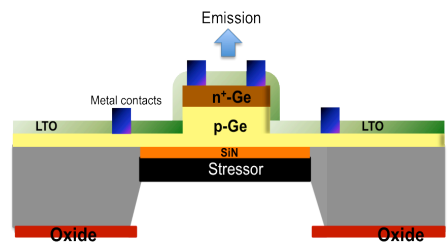

Fig. 1. Schematic of a strained Ge membrane LED

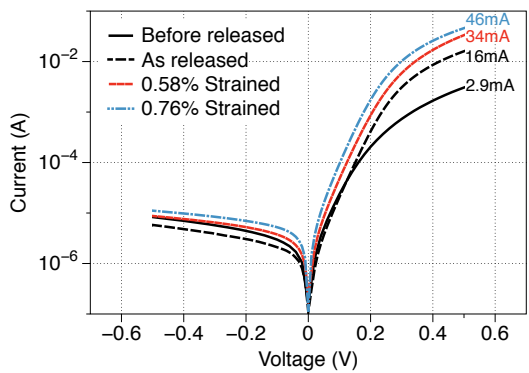

Fig. 2. (Color online) I-V characteristics of LEDs on several highly strained membranes. On-off ratio increases by one order of magnitude as Si is etched away. Strain further enhances both on and off current.

As shown in Fig. 1, 1.2μm thick p-Ge and 250nm thick n+-Ge were epitaxially grown on a Si substrate with a patterned layer of silicon dioxide on the backside. Then, a Ge mesa with a diameter of 200μm was defined by lithography and a 30nm thick low temperature oxide (LTO) was deposited as a passivation layer, followed by Ti/Ni metal contact deposition. In order to protect the top surface during the Si etch in TMAH, we used a commercially available protective coating called Protek B3. After the Si substrate was released and a Ge membrane was formed, a 20nm silicon nitride ($Si_3N_4$) was then deposited from the backside as a passivation layer and, finally, a tungsten stressor layer with approximately 4Gpa compressive stress was deposited to induce a large tensile strain in Ge active medium.

We performed electrical measurements at room temperature before and after etching away the Si. The solid black curve in Fig. 2 shows the I-V characteristic of a Ge PN diode on a Si substrate, showing an on-off ratio of approximately two orders of magnitude. After the Si is etched, the on-off ratio increases by one order of magnitude, as shown in the dotted black curve, due to a decrease in dark current and an increase in forward current. We believe that the dark current reduction is because etching the Si also removes some portion of the defective region at Si/Ge interface, causing a decrease in the recombination current under reverse bias. Additionally, in the absence of Si, the tighter electric field distribution within the thin Ge may extend the depletion region closer to the p-type metal contact, resulting in a smaller series resistance, and this may explain why the forward current under the bias of 0.5V increases almost by one order of magnitude. This forward current can be further enhanced with tensile strain because the band gap reduction increases the intrinsic carrier concentration, and possibly because electron mobility increases with tensile strain [14]. Since strain can increase the fraction of electrons in the direct Γ valley, which has a smaller effective mass than the indirect L valley, the overall electron mobility increases with strain. Also, because of the increased number of holes in the light-hole band due to the strain induced heavy-hole / light-hole energy splitting, the effective hole mobility increases, too. While an as-released LED shows a forward current of 16mA at 0.5V, the forward currents in 0.58% strained and 0.76% strained LEDs increase to 34mA and 46mA, respectively. The forward current enhancement will be much more prominent if we employ a P+-N diode where the hole current dominates because the hole mobility increases much faster than the electron mobility according to [15]. Direct measurements of mobility enhancement with strain are currently being investigated.

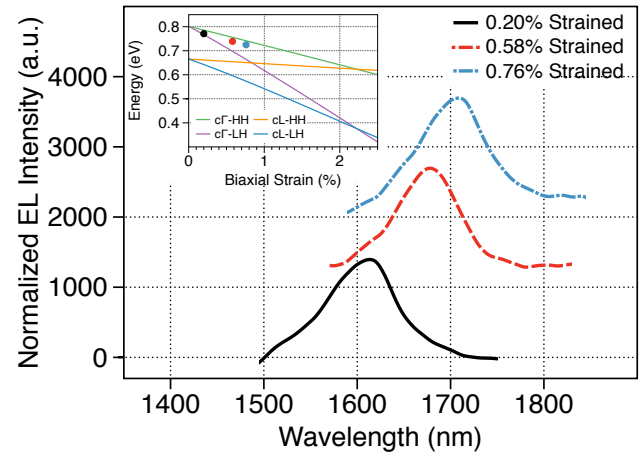

Fig. 3. (Color online) Normalized EL spectra from as-released, 0.58%, and 0.76% strained LEDs. A 0.76% strained LED shows a 100nm redshift of the center wavelength. The inset of Fig. 3 shows that the emission energies from these three samples are in good agreement with the calculated band gap energies.

EL measurements show successful light emission from strained Ge membrane LEDs. The measurements were conducted at room temperature at a current density of approximately 250A/$cm^2$. The emitted light was focused using a Mitutoyo 10x NIR objective lens with a numerical aperture (NA) of 0.26, and a strained InGaAs detector cooled to -100°C by liquid nitrogen was used to measure the signal over an extended wavelength range. As shown in Fig. 3, an as-released Ge LED emits a spectrum with a center wavelength of 1610nm, due to the 0.2% strain from thermal mismatch. By introducing tensile strains of 0.58% and 0.76% in the active region of PN diodes, we observed redshifts of the center wavelengths by 70nm and 100nm, respectively. These three spectra were normalized and the absolute EL intensities were not compared because the sample curvature in the highly strained LEDs can significantly affect the light collection efficiency, especially with an objective lens of such a small NA. According to previous simulations for the fraction of electrons in the direct Γ valley vs. strain [13], we expect to see an increase in the integrated EL intensity by a factor of 1.7 from 0.76% strained Ge LEDs compared to 0.2% strained cases once we factor in the collection efficiency. The inset of Fig. 3 shows that the emission energies from these three samples are in good agreement with the calculated band gap energies. When it comes to laser applications, strain effect on efficiency improvements will be much more significant than for LEDs because the optical net gain in a highly strained Ge is mainly contributed by the transition from the direct Γ valley to the light-hole band, and the smaller density of states (DOS) in the light-hole band facilitates population inversion, thus reducing the lasing threshold even further [16]. In the next section, therefore, we propose a VCSEL on a highly strained Ge membrane and discuss the extent to which strain reduces lasing threshold.

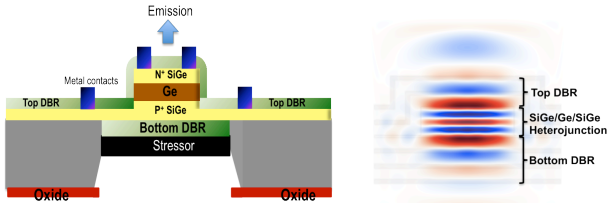

Fig. 4(a) Proposed VCSEL structure using a highly strained Ge as the gain medium. (b) FDTD simulations show that the optical mode can be tightly confined in the heterostructure.

Fig. 4(a) presents our proposed VCSEL structure using a highly strained Ge membrane as the gain medium. This design can be easily integrated by replacing the Ge PN layer in our Ge LEDs with a $p^+$-$Si_xGe_{1-x}$/n-Ge/$n^+$-$Si_xGe_{1-x}$ double heterojunction, and replacing the LTO with quarter wavelength $Si/SiO_2$ distributed Bragg reflectors (DBR). In this structure, moderately doped n-type Ge ($10^{18}cm^{-3}$) sandwiched between SiGe layers serves as the laser's gain region. With the Si composition x=0.15 in SiGe layers, we expect most injected carriers to be confined in the Ge region due to the type I band offset of the SiGe/Ge/SiGe interface [3]. While the bottom DBR layer consists of three alternating quarter-wave $Si/SiO_2$ layers with reflectance > 99%, the top DBR layer has only two alternating layers in order to preferentially emit light from the top surface. Using the transfer matrix method (TMM) and a tight-binding model, the thicknesses of SiGe barriers and of Ge gain medium are carefully designed to support a cavity mode at the wavelength of highest optical net gain. To discuss the strain effect on the lasing threshold, we compared two cases: 0.2% and 1.0% tensile strain in the Ge gain medium. Resonant wavelengths were chosen to be 1515nm and 1874nm for 0.2% and 1.0% strain cases, respectively. The Ge thickness was set at 348nm for 0.2% strain and 453nm for 1.0% strain in order to have one full wavelength within the gain medium. Using FDTD simulations in Fig. 4(b), it is shown that the optical mode can be tightly confined within SiGe/Ge/SiGe region due to the highly reflective DBRs. We find Q-factors of 677 and 598 for the 0.2% and 1% strain cases, respectively, and optical losses for the two cases are calculated to be $262cm^{-1}$ and $235cm^{-1}$, respectively. In order to achieve lasing in these structures, net modal gain (material gain divided by optical confinement factor) should exceed the optical loss from the cavity. With an optical confinement factor of approximately 0.3 in both cases that is calculated from TMM, the threshold gain coefficients for two cases are $873cm^{-1}$ and $783cm^{-1}$.

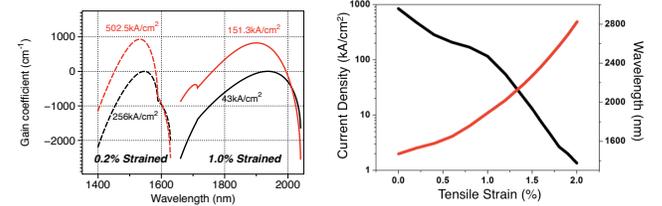

Fig. 5 (a) Gain spectra for 0.2% and 1% strained Ge cases, showing a 1% strained case reduces the threshold current density more than a factor of three compared to a 0.2% case. (b) Threshold current density and the wavelength of highest optical net gain vs. strain

Using a tight-binding model, and extrapolating the presumed absorption coefficients of [16] to higher strains, we obtained optical net gain spectra as a function of injection current density for both 0.2% and 1% strained Ge gain media cases as shown in Fig. 5(a). Note that our simulations assumed a crossover of the direct bandgap at 2.5% strain, whereas most sources claim the crossover occurs around 1.7- 2.0% [17-18], indicating that our results are a very conservative estimate. Because tensile strain lowers the direct Γ valley faster than the indirect L valley, and also because bi-axial tensile strain reduces the density of states (DOS) at the top of the valence band by light-hole / heavy-hole splitting, we find that increasing the strain from 0.2% to 1.0% reduces the required current density for positive optical net gain from $256kA/cm^2$ to $43kA/cm^2$. For our proposed structure, moving from 0.2% to 1% strain in the gain media reduces the threshold current density for lasing from $503kA/cm^2$ to $151kA/cm^2$. Particularly for room temperature laser operation, the reduction of threshold current density has a significant meaning because high currents can lead to serious heating problems, which may adversely affect laser operation. Fig. 5(b) shows the projection of threshold current density vs. strain up to 2% in a moderately n-type doped Ge ($10^{18}cm^{-3}$) gain medium, assumed threshold gain coefficient of $1000cm^{-1}$ is required to overcome cavity losses. While 1% strain reduces the threshold by approximately one order of magnitude compared to the unstrained case, pushing the strain up to 2% for the direct band gap Ge is expected to reduce the threshold current density by a factor of 1000 compared to the unstrained case. Since the wavelength of highest net gain would be red-shifted beyond 2μm, our tunable strained membrane technique can also be utilized for highly efficient mid-infrared laser applications.

In conclusion, we have fabricated highly strained Ge membrane LEDs and successfully demonstrated electroluminescence. A 0.76% strained LED showed a 100nm redshift in the spectrum and improved I-V characteristics. As simulations show that strain can significantly reduce the threshold current density for lasing, we believe that our strained membrane technique can play a crucial role in achieving an efficient Ge laser for on-chip optical interconnects.

**Acknowledgements**

This work was supported by U.S. Government through APIC (Advanced Photonic Integrated Circuits) corporation and the Connectivity Center of the Focus Center Research Program, a National Science Foundation Graduate Research Fellowship under Grant No. DGE-0645962, a Stanford Graduate Fellowship and a Fonds Québécois de la Recherche sur la Nature et les Technologies Master's Research Scholarship.